\documentclass{PoS}

\usepackage{cite}

\newcommand{\bd}{\begin{displaymath}}
\newcommand{\ed}{\end{displaymath}}

\title{Leading order hadronic contribution to g-2 from twisted mass QCD}

\ShortTitle{Leading order hadronic contribution to g-2 from twisted mass QCD}

\author{\speaker{Dru B.\ Renner}\thanks{Current address:\ Jefferson Lab.}\,\,$^a$,
Xu Feng\thanks{Current address:\ KEK.}\,\,$^{a, b}$,
Karl Jansen$^{a}$ and
Marcus Petschlies$^{c}$\\
\llap{$^a$}NIC, DESY, Platanenalle 6, D-15738 Zeuthen, Germany\\
\llap{$^b$}Universit\"at M\"unster, Institut f\"ur Theoretische Physik, Wilhelm-Klemm-Strasse 9, D-48149 M\"unster, Germany\\
\llap{$^c$}Institut f\"ur Elementarteilchenphysik, Fachbereich Physik, Humboldt Universit\"at zu Berlin, D-12489, Berlin, Germany
}

\abstract{We calculate the leading order hadronic contribution to the
  muon anomalous magnetic moment using twisted mass lattice QCD.  The
  pion masses range from $330~\mathrm{MeV}$ to $650~\mathrm{MeV}$.  We
  use two lattice spacings, $a=0.079~\mathrm{fm}$ and
  $0.063~\mathrm{fm}$, to study lattice artifacts.  Finite-size
  effects are studied for two values of the pion mass, and we
  calculate the disconnected contributions for four ensembles.
  Particular attention is paid to the dominant contributions of the
  vector mesons, both phenomenologically and from our lattice
  calculation.
\vspace{15pt}
\begin{center}
\includegraphics[width=100pt]{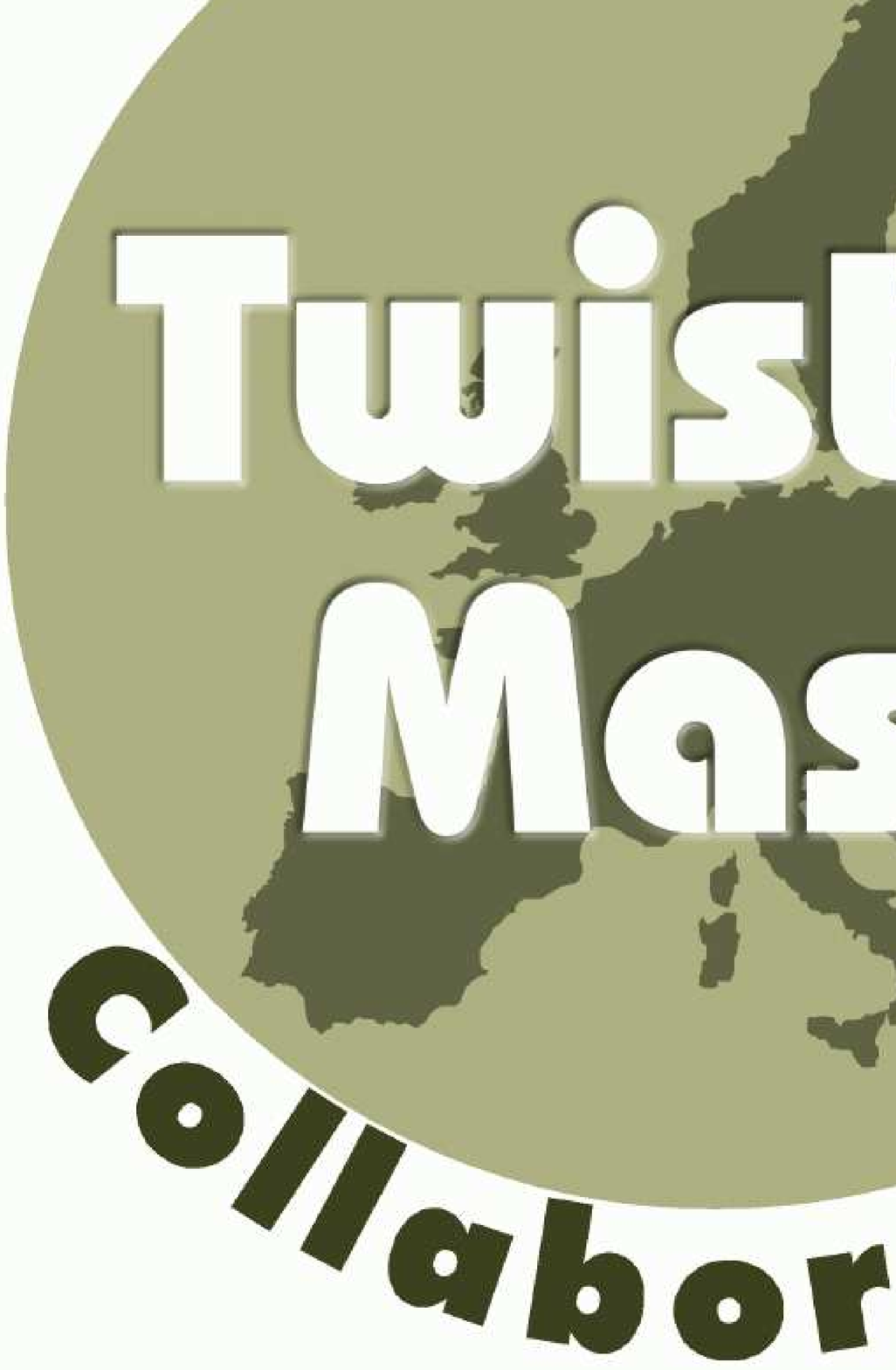}
\end{center}
  }

\FullConference{The XXVIII International Symposium on Lattice Field Theory\\
June 14-19,2010\\
Villasimius, Sardinia Italy}

\begin{document}

\section{Introduction}

A persistent discrepancy of about $3.2\,\sigma$ between the
measured~\cite{Bennett:2006fi} and the theoretically
calculated~\cite{Jegerlehner:2009ry} anomalous magnetic moment of the
muon, $a_\mu$, has motivated several lattice calculations of the
leading order hadronic contribution, $\Delta\,
a_\mu$~\cite{Blum:2002ii,Gockeler:2003cw,Aubin:2006xv,Renner:2009by,Brandt:2010ed}.
This contribution is the dominant source of error in the theoretical
estimate of $a_\mu$ and presents both a challenge and opportunity for
lattice QCD calculations.  Here we extend our previous
work~\cite{Renner:2009by} to address most of the systematic errors in
an effort to establish what is required for a precise calculation of
$\Delta\,a_\mu$.  A detailed analysis will be presented elsewhere, so
these proceedings will focus on the contributions of the vector
mesons.
% to $\Delta\,a_\mu$.

\section{Leading order hadronic contribution to $a_\mu$}

The leading order hadronic contribution to the anomalous magnetic
moment of the muon can be calculated from the vacuum polarization
tensor
%with Euclidean momentum
using the following expression from~\cite{Blum:2002ii}.
\begin{equation}
\label{al}
\Delta\, a_\mu = \alpha^2 \int_0^{\infty}\!\! dQ^2\,\, \frac{1}{Q^2} w(Q^2/m_l^2) \Pi_R(Q^2)
\end{equation}
The renormalized vacuum polarization function is $\Pi_R(Q^2) =
\Pi(Q^2) - \Pi(0)$ and the weight function is given
in~\cite{Blum:2002ii}.
%
%\begin{displaymath}
%\Pi_R(Q^2) = \Pi(Q^2) - \Pi(0)
%\end{displaymath}
%
%and the weight function, written was a function of $r=Q^2/m_l^2$, is
%
%\begin{displaymath}
%w(r) = \frac{64}{r^2 ( 1 + \sqrt{ 1 + 4/r } )^4 \sqrt{ 1 + 4/r } }\,.
%\end{displaymath}
%
%Making the change of variables $r=x^2/(1-x)$, we can recast
%Eq.~\ref{al} as
%%
%\begin{equation}
%\label{al2}
%\Delta\, a_l = 4 \alpha^2 \int_0^1\!\! dx\,\, (1-x) \Pi_R(x^2 m_l^2/(1-x))\,.
%\end{equation}
%
The relevant lattice details for this calculation can be found in
\cite{Renner:2009by}.

%\subsection{Phenomenological estimates of $\Delta\,a_\mu$}
%
The hadronic correction to $a_\mu$ is normally estimated using a
dispersion relation and experimental measurements of the cross section
$\sigma(e^+e^-\rightarrow \mathrm{hadrons})$.  A recent review
\cite{Jegerlehner:2009ry} provides a list of results ranging from
$\Delta\,a_\mu=684.6~(6.4) \cdot 10^{-10}$ to $711.0~(5.8) \cdot
10^{-10}$.  The variation in the experimental estimates is beyond our
current precision, so in the following we will use just the estimate
of $691.0~(5.3)\cdot 10^{-10}$ from \cite{Jegerlehner:2008rs}.  More
important than the total value for $\Delta\, a_\mu$ are the individual
contributions.  The low energy region is dominated by the vector
mesons:\ $\rho$, $\omega$ and $\phi$.  The contributions from each are
given in \cite{Jegerlehner:2008rs} as
\begin{displaymath}
\left. \Delta\, a_\mu \right|_\rho = 501.1~(3.0) \cdot 10^{-10}
~~~~
\left. \Delta\, a_\mu \right|_\omega = 37.0~(1.1) \cdot 10^{-10}
~~~~
\left. \Delta\, a_\mu \right|_\phi = 34.42~(0.9) \cdot 10^{-10}\,.
\end{displaymath}
Taken together, these three contributions already account for about
$83\%$ of the total $\Delta\,a_\mu$.  Verifying the contributions from
the vector mesons is clearly a good first step towards calculating the
complete $\Delta\,a_\mu$.

\section{Vector mesons}

As discussed in previous lattice calculations of $\Delta\,a_\mu$, the
low $Q^2$ region of $\Pi_R(Q^2)$ requires an interpolation and
extrapolation in order to perform the integration in Eq.~\ref{al}.
Currently this requires introducing some model assumptions.  Given the
significant contributions of the lightest vector mesons, it is
reasonable to incorporate their contributions into the $Q^2$ shape of
the model.  First we establish what is known regarding the
electromagnetic decays of the $\rho$, $\omega$ and $\phi$.  Then we
discuss the additional model assumptions needed to calculate the
contributions of the vector mesons to $\Delta\,a_\mu$.

\subsection{Vector meson decay constants}

The electromagnetic coupling of a vector meson is defined by
\begin{displaymath}
\langle \Omega | J^{em}_\mu(0) | V, p, \epsilon \rangle = m_V^2 g_{V,em} \epsilon_\mu(p)
~~~~~~~~
f_{V,em} = m_V g_{V,em}\,.
\end{displaymath}
%
%and it is common to form a dimensionful decay constant as
%%
%\begin{displaymath}
%f_{V,em} = m_V g_{V,em}\,.
%\end{displaymath}
%
The value of $g_{V,em}$ can be determined through the decay
$V\rightarrow e^+ e^-$.  A straightforward exercise gives the partial
width for this decay as
\begin{displaymath}
\Gamma(V\rightarrow e^+ e^-) = \frac{4\pi}{3} \alpha_{em}^2 g^2_{V,em} m_V ( 1 + 2 m_e^2 / m_V^2) \sqrt{ 1 - 4 m_e^2 / m_V^2 }\,.
\end{displaymath}
Using the latest values from the PDG~\cite{PDG}, we find
\begin{displaymath}
g_{\rho,em} = 0.20174\,(86)
~~~~
g_{\omega,em} = 0.05863\,(98)
~~~~
g_{\phi,em} = 0.07473\,(120)\,.
\end{displaymath}
It is common to represent these decay constants in an isospin basis
defined as
\begin{displaymath}
J^{I=0}_\mu=(\overline{u}\gamma_\mu u + \overline{d}\gamma_\mu d)/\sqrt{2}
~~~~
J^{I=1}_\mu=(\overline{u}\gamma_\mu u - \overline{d}\gamma_\mu d)/\sqrt{2}
~~~~
J^s_\mu = \overline{s}\gamma_\mu s\,.
\end{displaymath}
Assuming $m_u=m_d$, $\phi$ is a pure $\overline{s} s$ state and
$\omega$ has no $\overline{s} s$ contribution yields
\begin{displaymath}
f_\rho = \sqrt{2} f_{\rho,em}
~~~~
f_\omega = 3 \sqrt{2} f_{\omega,em}
~~~~
f_\phi = -3 f_{\phi,em}\,
\end{displaymath}
where $\langle \Omega | J_\mu | V, p, \epsilon \rangle = m_V f_V
\epsilon_\mu(p)$ defines the isospin projected decay constants and the
current is understood to correspond to the given meson.  The resulting
numerical values are
\begin{displaymath}
f_\rho = 221.3\,(1.0)~\mathrm{MeV}
~~~~
f_\omega = 194.7\,(3.2)~\mathrm{MeV}
~~~~
f_\phi = 228.7\,(3.6)~\mathrm{MeV}\,,
\end{displaymath}
which are all of the same order of magnitude.  This is the expectation
from chiral perturbation theory assuming precisely this pattern of
$\omega$-$\phi$ mixing.  Letting $f_V$ denote the generic vector meson
decay constant in this case gives
\begin{displaymath}
f_{\rho,em} = \frac{f_V}{\sqrt{2}}
~~~~
f_{\omega,em} = \frac{f_V}{3\sqrt{2}}
~~~~
f_{\phi,em} = -\frac{f_V}{3}\,.
\end{displaymath}
This can be generalized to include the $\omega$-$\phi$ mixing with an
angle $\theta$, resulting in
\begin{displaymath}
f_{\rho,em} = \frac{f_V}{\sqrt{2}}
~~~~
f_{\omega,em} = \sin\theta \frac{f_V}{\sqrt{6}}
~~~~
f_{\phi,em} = -\cos\theta \frac{f_V}{\sqrt{6}}\,.
\end{displaymath}
As needed shortly, we note that $\tan\theta = 1/\sqrt{2}$ is the angle that
corresponds to a pure $\overline{s} s$ state for the $\phi$ and this angle reduces
the previous line to the line before.
%We note that $\tan\theta = 1/\sqrt{2}$ is the angle that
%corresponds to a pure $\overline{s} s$ state for the $\phi$.

\subsection{Vector meson contributions to $\Delta\,a_\mu$}

%The previous discussion relating the electromagnetic couplings of the
%vector mesons to their $e^+ e^-$ decays is model independent, but 
%the relationships between the couplings is model dependent.  Additionally we
%want to determine the contribution of the mesons to $\Delta\,a_\mu$.

The couplings in the previous section are on-shell properties of the
mesons, but now we must discuss the off-shell aspects of the vectors.
This inevitably introduces a model dependence.  Tree-level
calculations in chiral perturbation theory provide a definite
off-shell form of the vector meson propagator and give a contribution
to the renormalized vacuum polarization function of
\begin{displaymath}
\Pi_{R,V}(Q^2) = \frac{f_{V,em}^2}{m_V^2} \frac{Q^2}{Q^2+m_V^2}\,.
\end{displaymath}
This expression can also be achieved by simply assuming an off-shell
propagator of the form $A/(Q^2+m_V^2)$ and fixing $A$ by demanding the
correct $\Gamma(V\rightarrow e^+e^-)$ result.

Combining the above result with the relationships between the decay
constants gives
\begin{displaymath}
\Pi_{R,\rho+\omega+\phi}(Q^2) = f_V^2\left\{ \frac{1}{2} \frac{Q^2/m_\rho^2}{Q^2+m_\rho^2}
+ \frac{\sin^2\theta}{6} \frac{Q^2/m_\omega^2}{Q^2+m_\omega^2}
+ \frac{\cos^2\theta}{6} \frac{Q^2/m_\phi^2}{Q^2+m_\phi^2} \right\}\,.
\end{displaymath}
Additionally setting $m_\rho = m_\omega = m_\phi = m_V$ gives
\begin{displaymath}
\Pi_{N_f=3}(Q^2) = \frac{2f_V^2}{3m_V^2} \frac{Q^2}{Q^2+m_V^2}\,.
\end{displaymath}
In the above we now make explicit that this is envisioned as a
three-flavor result.  We can also extract the two-flavor result by
decoupling the $\phi$ and demanding a pure $\overline{s} s$ state for
the $\phi$ by setting $\tan\theta = 1/\sqrt{2}$.  The resulting
expression is
\begin{displaymath}
\Pi_{N_f=2}(Q^2) = \frac{5f_V^2}{9m_V^2} \frac{Q^2}{Q^2+m_V^2}\,.
\end{displaymath}
Notice that the strength of the $N_f = 2$ and $3$ results follows the
sum of the charges squared.

The resulting integral in Eq.~\ref{al} can be performed giving the
following contributions for each of the vector mesons.
\begin{displaymath}
\Delta\, a_{\mu,\rho} = 470.9.~(4.0) \cdot 10^{-10}
~~~~
\Delta\, a_{\mu,\omega} = 39.1~(1.3) \cdot 10^{-10}
~~~
\Delta\, a_{\mu,\phi} = 39.0~(1.3) \cdot 10^{-10}\,.
\end{displaymath}
These values already reproduce much of the experimentally determined
contributions given earlier.  The discrepancy is largest for the
$\rho$ and is a likely indicator of the effects of the
$\rho\rightarrow\pi\pi$ decay.

\section{Lattice calculation}

\begin{table}
\begin{center}
\begin{tabular}{|c|c|c|c|c|c|c|c|} \hline
$\beta$ & $L/a$ & $a\mu$ & $m_\pi~[\mathrm{MeV}]$ & $m_V~[\mathrm{MeV}]$ & $f_V~[\mathrm{MeV}]$ & $\Delta\,a_{\mu,V}~[10^{-10}]$ \\\hline
3.90 & 20 & 0.0040 & 347.9~(6.2) & 1209.~(161.) & 345.~(64.) & 209.~(45.) \\\hline
3.90 & 24 & 0.0150 & 645.9~(1.7) & 1235.~(23.) & 322.~(10.) & 164.~(6.)\,\,\, \\\hline
3.90 & 24 & 0.0100 & 524.5~(1.2) & 1156.~(36.) & 312.~(13.) & 199.~(11.) \\\hline
3.90 & 24 & 0.0085 & 484.6~(1.2) & 1144.~(34.) & 305.~(13.) & 198.~(9.)\,\,\, \\\hline
3.90 & 24 & 0.0064 & 423.1~(1.0) & 1083.~(35.) & 297.~(11.) & 234.~(15.) \\\hline
3.90 & 24 & 0.0040 & 340.2~(1.7) & 1067.~(46.) & 290.~(15.) & 237.~(19.) \\\hline
3.90 & 32 & 0.0040 & 334.2~(0.5) & 1044.~(51.) & 296.~(17.) & 267.~(26.) \\\hline
3.90 & 32 & 0.0030 & 291.5~(1.0) & 956.~(70.) & 260.~(20.) & 295.~(45.) \\\hline
4.05 & 24 & 0.0060 & 453.5~(3.4) & 1094.~(39.) & 308.~(14.) & 241.~(15.) \\\hline
4.05 & 32 & 0.0080 & 517.1~(1.6) & 1124.~(36.) & 298.~(12.) & 204.~(13.) \\\hline
4.05 & 32 & 0.0060 & 448.5~(1.9) & 1104.~(45.) & 299.~(17.) & 219.~(14.) \\\hline
4.05 & 32 & 0.0030 & 325.1~(1.9) & 1063.~(78.) & 297.~(25.) & 254.~(35.) \\\hline
\end{tabular}
\end{center}
\caption{
  Ensembles and results used in this work.
  %The bare coupling is given by
  %$g^2=6/\beta$.  The bare quark mass in lattice units is $a\mu$,
  %and the lattice volume in lattice units is $(L/a)^3\times (T/a)$ with $T = 2L$.  
  The pion
  masses are taken from \cite{Baron:2009wt} with one exception.  The $L/a=20$ result
  was communicated privately by the ETMC collaboration.  The lattice spacings 
  are $a=0.079\,(2)~\mathrm{fm}$ and $0.063\,(1)~\mathrm{fm}$
  for $\beta=3.90$ and $4.05$~\cite{Baron:2009wt}.  Results from this work are
  the vector meson masses $m_V$, decay constants $f_V$ and the corresponding
  contribution to $\Delta\,a_\mu$ (the last two are normalized as the $\rho$ contribution).  Up to
  disconnected contributions, the $N_f=2$ value of 
  $\Delta\,a_\mu$ is $10/9$ of $\Delta\,a_{\mu,V}$.  Additionally ignoring $s$ quark quenching effects, the
  $N_f=3$ value is $4/3$ of $\Delta\,a_{\mu,V}$.}
\label{en}
\end{table}
The vector decay constant can be calculated from the correlator of the
non-singlet current as follows.
\begin{displaymath}
\sum_i \int\!\! d^3\vec{x}\,\, \langle J^{I=1}_i(t,\vec{x}) J^{I=1}_i(0) \rangle
\rightarrow
\frac{ e^{-m_V t } }{ 2 m_V } \sum_{\vec{\epsilon}} \langle \Omega | J_i | V, \vec{\epsilon} \rangle 
\langle V, \vec{\epsilon} | J_i | \Omega \rangle
= \frac{ 3 m_V f_V^2 }{2} e^{-m_V t}
\end{displaymath}
This correlator can be calculated on the lattice without disconnected
diagrams.  Let $C_{\mu\nu}(x)$ denote the connected piece of the
single quark correlator $\langle J^q_\mu(x) J^q_\nu(0)\rangle$.
Similarly let $D_{\mu\nu}(x)$ be the disconnected piece.  Then the
$I=1$ correlator is
\bd
\langle J^{I=1}_\mu(x) J^{I=1}_\nu(0) \rangle = C_{\mu\nu}(x)\,.
\ed
Additionally, the electromagnetic coupling of the $\rho$ can also be
calculated directly without disconnected diagrams.
\bd
\langle J^{em}_\mu(x) J^{I=1}_\nu(0) \rangle = \frac{1}{\sqrt{2}} C_{\mu\nu}(x)
\ed
This is consistent with the earlier relationship $f_{\rho,em} =
f_\rho/\sqrt{2}$.

The $\omega$ and $\phi$ decay constants are then obtained from
\bd
\langle J^{I=0}_\mu(x) J^{I=0}_\nu(0) \rangle = C_{\mu\nu}(x) + 2 D_{\mu\nu}(x)
\ed
and
\bd
\langle J^{s}_\mu(x) J^{s}_\nu(0) \rangle = C_{\mu\nu}(x) + D_{\mu\nu}(x)\,.
\ed
In the last expression we are assuming a quenched strange quark that
is degenerate with the $u$ and $d$ quarks.  Thus ignoring disconnected
diagrams we find $m_\omega = m_\phi = m_\rho$ and $f_\omega = f_\phi =
f_\rho$.  The electromagnetic couplings can also be calculated using
the following expressions.
\bd
\langle J^{em}_\mu(x) J^{I=0}_\nu(0) \rangle = \frac{1}{3\sqrt{2}} C_{\mu\nu}(x) + \frac{5}{3\sqrt{2}}D_{\mu\nu}(x)
\ed
\bd
\langle J^{em}_\mu(x) J^{s}_\nu(0) \rangle = -\frac{1}{3} C_{\mu\nu}(x)
\ed
Again we see that up to $s$ quark quenching effects and disconnected
diagrams the results $f_{\omega,em} = f_\omega/(3\sqrt{2})$ and
$f_{\phi,em} = -f_\phi / 3$ are reproduced.  This makes it clear that
the theory with a quenched strange quark should be compared to the
$\tan\theta = 1/\sqrt{2}$ scenario in which the $\phi$ is pure
$\overline{s}s$ state.

The full electromagnetic vacuum polarization function in the $N_f=2$
calculation is
\bd
\left. \langle J^{em}_\mu(x) J^{em}_\nu(0) \rangle \right|_{N_f=2} = \frac{5}{9} C_{\mu\nu}(x) + \frac{1}{9} D_{\mu\nu}(x)
\ed
and the result including a quenched and degenerate strange quark is
\bd
\left. \langle J^{em}_\mu(x) J^{em}_\nu(0) \rangle \right|_{N_f=3} = \frac{2}{3} C_{\mu\nu}(x)\,.
\ed
Again notice that, up to disconnected contributions and quenching
effects, the $N_f=2$ and $3$ results agree to within the sum of
charges squared.

\section{Results}

Our earlier calculation~\cite{Renner:2009by} has been extended to
include a second lattice spacing, more quark masses, volumes and
estimates of disconnected diagrams.  (See Tab.~\ref{en}.)
%A detailed analysis will be presented in a forthcoming publication.  
Here we focus on the simplest model for $\Pi_R(Q^2)$ that includes
only the vector mesons.  The vector meson masses and decay constants
and the corresponding contribution to $\Delta\,a_\mu$ are given in
Tab.~\ref{en}.  The vector contribution to $\Delta\,a_\mu$, normalized
as the $\rho$, rises from $164~(6) \cdot 10^{-10}$ to $295~(45) \cdot
10^{-10}$ but falls well short of the expected $501.1~(3.0) \cdot
10^{-10}$.

\begin{figure}
\begin{minipage}{210pt}
\includegraphics[width=210pt]{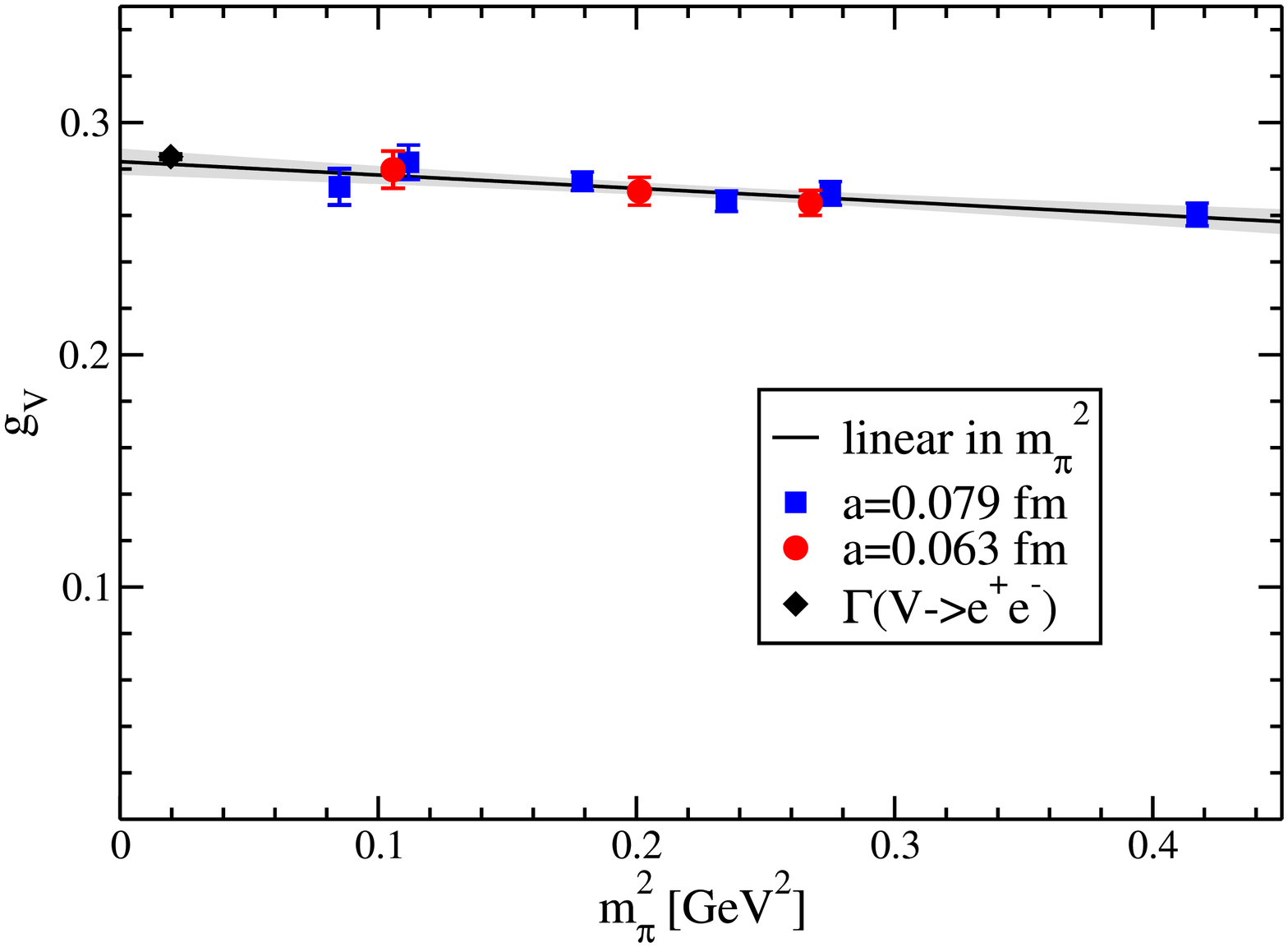}
\caption{Vector coupling.  The vector coupling, normalized as the
  $\rho$, calculated for $a=0.079~\mathrm{fm}$ and
  $a=0.063~\mathrm{fm}$ agree.  A linear extrapolation in $m_\pi^2$ to
  the physical pion mass agrees with the value extracted from
  $\Gamma(V\rightarrow e^+ e^-)$.  The factor of $m_V$ in $f_V$ hides
  this agreement.}
\label{gv}
\end{minipage}
\hspace{4pt}
\begin{minipage}{210pt}
\includegraphics[width=210pt]{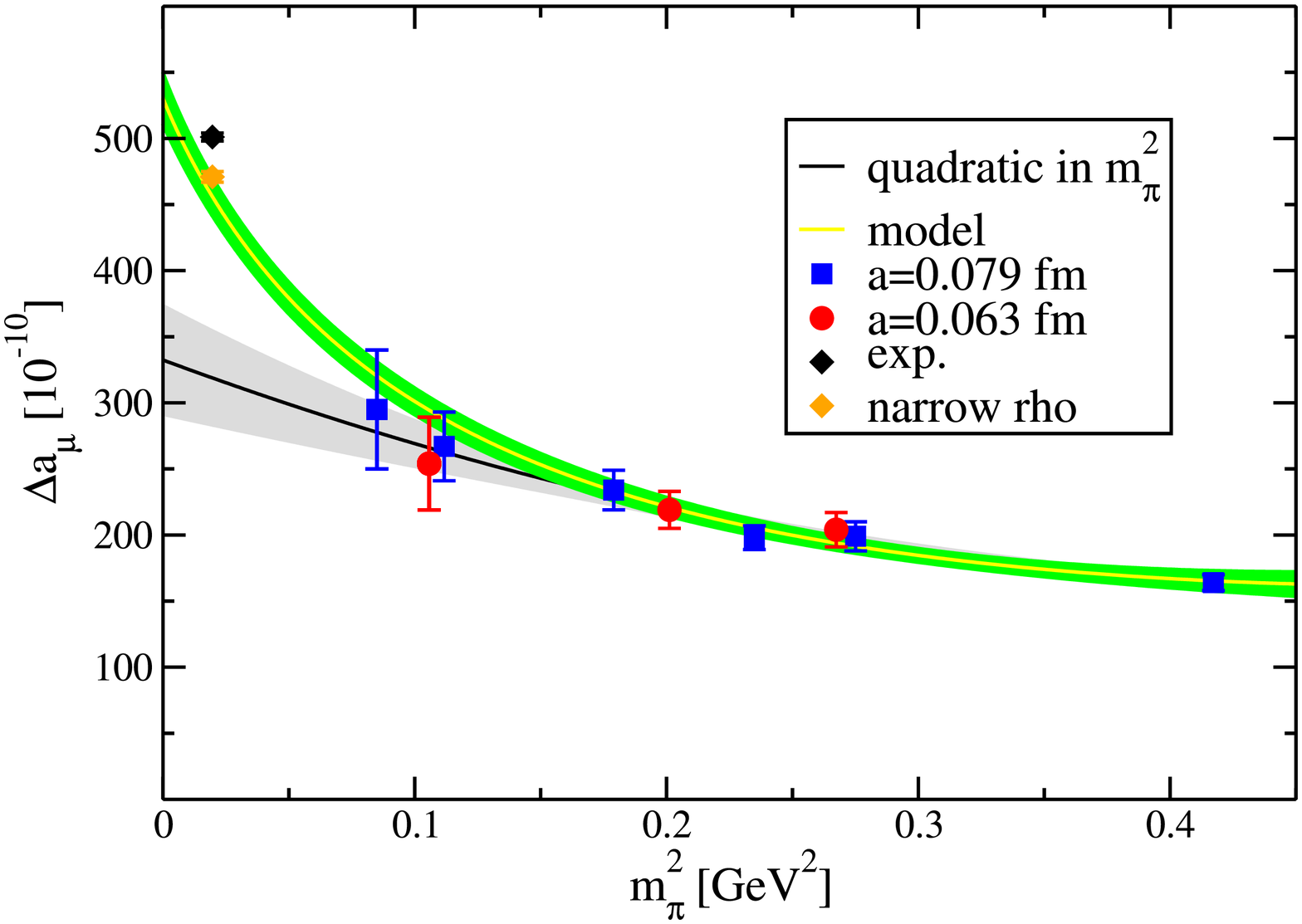}
\caption{Vector contribution to $\Delta\,a_\mu$.  A quadratic
  extrapolation in $m_\pi^2$ undershoots the measured contribution of
  the $\rho$ significantly.  A model (described in text) suggests the
  importance of calculating at small enough $m_\pi$ for the vector
  mass to approach the physical $\rho$ mass.}
\label{amuv}
\end{minipage}
\end{figure}

We suspect that this discrepancy is mainly due to the large value of
the $\rho$ mass found for the range of $\pi$ masses in Tab.~\ref{en}.
To illustrate this point, we plot the coupling of the vector meson in
Fig.~\ref{gv}.
There we see that $g_{V}$ appears to have a mild quark mass dependence
and a simple linear extrapolation in $m_\pi^2$ gives a value in
agreement with the experimental measurement.  We can isolate the
dependence of $\Delta\,a_{\mu,V}$ on $g_{V}$ and $m_V$ by integrating
Eq.~\ref{al} exactly and then expanding in $m_\mu/m_V$.  The first few
terms are
\bd
\Delta\,a_{\mu,V} = 2 \alpha^2 g_{V}^2 \left(\frac{m_\mu}{m_V}\right)^2\left\{ \frac{1}{3} + \frac{25}{12}\left(\frac{m_\mu}{m_V}\right)^2 + \left(\frac{m_\mu}{m_V}\right)^2 \ln \left(\frac{m_\mu}{m_V}\right)^2 + {\cal O} \left(\frac{m_\mu}{m_V}\right)^4 \right\}\,.
\ed
This makes it plausible that the discrepancy is simply due to the
large value of the $\rho$ mass and the approach to the physical point
should account for most of the discrepancy between the current values
of $\Delta\,a_{\mu,V}$ and the experimental measurement.  To
illustrate this, we fit the chiral expansion $m_V = a + b m_\pi^2 + c
m_\pi^3$ to \emph{both} our lattice results and also the physical
value of the $\rho$ mass.  Combining this with the fit to $g_V$ we
produce the model curve in Fig.~\ref{amuv}.  Because $g_V$
extrapolates to the physical value and $m_V$ was constrained to give
the physical $\rho$ mass, this model automatically reproduces the
narrow $\rho$ approximation from earlier.

% This illustrates more clearly that the current values of
%$\Delta\,a_{\mu,V}$ and the experimental %result can be reconciled
%once that $\pi$ masses are light enough that the physical $\rho$ mass
%can be reproduced.

\section{Conclusions}

The final results of this work will be given in a forthcoming
publication.  Here we focus on understanding the dominant contribution
from the vector mesons.  Our current calculations produce values of
$\Delta\,a_\mu$ that are significantly lower than the experimental
measurements.  We construct a model that suggests that this effect may
be caused partially by a $\rho$ mass that is still rather large
compared to its physical value.
%Only once the $\pi$ mass is light enough that the $\rho$ mass
%approaches its physical value will $\Delta\,a_\mu$ take values near
%the experimental measurement.
Thus we expect that calculations with the physical value of the $\pi$
mass and including the strange and charm quark contributions will be
capable of achieving the precision necessary to match the accuracy of
current experimental measurements of $\Delta\,a_\mu$.

\section{Acknowledgments}

We thank Carsten Urbach for his valuable collaboration, and we thank
the John von Neumann Institute for Computing (NIC), the J{\"u}lich
Supercomputing Center and the DESY Zeuthen Computing Center for their
computing resources.  This work has been supported in part
by the DFG Sonder\-for\-schungs\-be\-reich/Transregio SFB/TR9-03 and
the DFG project Mu 757/13 and is coauthored in part by Jefferson 
Science Associates, LLC under U.S. DOE Contract No. DE-AC05-06OR23177.

\bibliographystyle{hunsrt}
\bibliography{lat10-renner.bib}

\end{document}